\documentclass[aps,pra,twocolumn]{revtex4-2}
\usepackage{color}
\usepackage{longtable}
\usepackage{dcolumn}

\newcolumntype{w}[1]{D{.}{.}{#1}}
\usepackage{graphicx}
\usepackage{bm}
\usepackage{bbm}
\usepackage{times}
\usepackage{nicefrac}
\usepackage{amsmath}
\usepackage{amsfonts}
\usepackage{amssymb}
\usepackage{amsthm}
\usepackage{physics}
\newcommand{\balpha}{\vec{\alpha}}

\newcommand{\bNabla}{\vec{\nabla}}
\newcommand{\vare}{\varepsilon}

\newcommand{\lbr}{\langle}
\newcommand{\rbr}{\rangle}

\newcommand{\Za}{Z\alpha}

\allowdisplaybreaks
\newcommand{\wpp}{\overline{w}}

\begin{document}

\title{Nuclear recoil effects with a realistic nuclear charge form factor}

\author{Vladimir A. Yerokhin}
\affiliation{Max-Planck-Institut f\"ur Kernphysik, Saupfercheckweg 1, 69117 Heidelberg, Germany}

\author{Tymon Kilich}
\affiliation{Faculty of Applied Physics and Mathematics, Gda\'nsk University of Technology, 80-233 Gda\'nsk, Poland}

\author{Krzysztof Pachucki}
\affiliation{Faculty of Physics, University of Warsaw,
             Pasteura 5, 02-093 Warsaw, Poland}

\date{\today}
\begin{abstract}

We introduce a novel two-parameter Gamma model of the nuclear charge distribution as an alternative to the widely used Fermi distribution. The model yields simple analytical formulas for all moments of the charge distribution and closed-form expressions for the electrostatic potential and momentum-space form factor, the latter satisfying the correct analyticity requirements for a physical charged particle. We apply the model to calculate the nuclear recoil correction to the Lamb shift in electronic and muonic atoms and analyze its nuclear-model dependence. We find that the nuclear-size recoil corrections to energy levels of heavy muonic atoms are strongly sensitive to higher moments of the nuclear charge distribution, highlighting the importance of realistic nuclear modeling. To facilitate further calculations, we also provide the Fortran code RECFNS, designed for computing the nuclear recoil correction
within the QED framework. The present work establishes a basis for future extensions to higher orders in the electron-to-nucleus mass ratio, hyperfine splitting, and radiative corrections.

\end{abstract}

\maketitle


\section{Introduction}

Atomic energy levels are affected by the finite charge distribution of the nucleus.
In the approximation of an infinitely heavy nucleus, the finite nuclear size (fns) effect can be
readily taken into account by using the appropriately modified Coulomb potential in the Dirac equation.
The fns energy shifts are approximately proportional to the mean-square charge radius of the nucleus,
$r_C^2$, which provides the basis for determining nuclear charge radii from atomic spectroscopy,
in particular from muonic atoms \cite{fricke:04}.

The theoretical treatment of fns effects becomes much more complicated for nuclei of finite mass.
In the relativistic theory, the dependence of atomic energy levels on the nuclear mass is highly nontrivial.
The well-known reduced mass can be meaningfully introduced only in the nonrelativistic limit, i.e., for the Schr\"odinger equation.
The use of the reduced mass in the Dirac equation -- even though it sometimes appears in the literature
\cite{indelicato:13, xie:23, yoshimura:25, liborio:26} -- is inconsistent
with relativistic theory. In fact, a proper treatment of nuclear recoil effects
requires a systematic QED treatment.

General expressions for the nuclear recoil effect to first order in the electron-nucleus mass ratio
$m/M$ and to all orders in the nuclear binding strength parameter $Z\alpha$ ($Z$ is the nuclear
charge number and $\alpha$ is the fine-structure constant) were obtained
for the point nucleus by Shabaev in
Refs.~\cite{shabaev:85,shabaev:88} (see also \cite{shabaev:98:rectheo}) and later rederived by one of
the present authors \cite{pachucki:95}.
A few years ago, the corresponding formulas were extended to include finite nuclear size effects
\cite{pachucki:23:prl} and higher orders in the mass ratio \cite{pachucki:24:hpqed}
within the framework of heavy-particle quantum electrodynamics (HPQED).

Treatment of the nuclear recoil effects for an extended nucleus is important for accurate predictions of energy levels in muonic atoms and, consequently, for reliable determinations of nuclear charge radii. For low-lying states of muonic atoms, the point-nucleus approximation is not applicable even as a starting approximation. Recently, it was demonstrated~\cite{beyer:26} that the long-standing  ``fine-structure anomaly'' in muonic $^{90}$Zr can be resolved by consistently
incorporating the nuclear recoil effect using the formulas derived in Ref.~\cite{pachucki:23:prl}. A rigorous treatment of nuclear recoil with a realistic
nuclear charge form factor is presented in this work.

One of the key features of the extended-nucleus nuclear-recoil formulas derived in Ref.~\cite{pachucki:23:prl}
and described later in this paper is that they require the nuclear form factor $F(k^2)$,
defined as a Fourier transform of the charge density $\rho(r)$,
\begin{align}\label{eqFF}
F(k^2) =&\ \frac{4\pi}{k}\int_0^{\infty}\rho(r)\,\sin(k\,r)\,rdr\,,
\end{align}
to be known not only for space-like  momentum transfer, $k^2>0$, but also for time-like momentum transfer $k^{2} <0$.
Consequently,
the nuclear charge form factor must have the correct analytic properties in the  complex $k^2$ plane.

It is known (see, e.g., Refs.~\cite{dumbrajs:80,alberto:21}) that the form factor of an arbitrary particle
is an analytic function on the complex $k^2$ plane,
vanishing at complex infinity, with a branch cut (or poles) on the
negative real axis, $-\infty < k^2  <0$.
On the cut, the form factor is related  to the creation and annihilation of real particles  and is
a complex quantity, whereas for $k^2 > 0 $ it describes elastic scattering and is real.

Surprisingly, it turns out that most (all but one) of the nuclear charge-distribution models commonly used in the literature do not satisfy the above-mentioned analyticity properties.
For example, the one-parameter Gaussian charge distribution
\begin{align}
\rho_\mathrm{1pG}(r) =&\ \frac1{(2\pi a)^{3/2}}\,e^{-\frac{r^2}{2a}}\,,
\end{align}
becomes in momentum space
\begin{align}
F_\mathrm{1pG}(k^2) =&\ e^{-\frac{a k^2}{2}}\,,
\end{align}
which exponentially grows for imaginary $k$.
Another commonly used model,
the one-parameter homogeneously charged sphere distribution
\begin{align}
\rho_\mathrm{1pH}(r) =&\ \frac{3}{4\,\pi\,R^3}\,\Theta(R-r)\,,
\end{align}
becomes in momentum space
\begin{align}
F_\mathrm{1pH}(k^2)=&\ \frac{3}{k\,R}\,j_1(k\,R)\,,
\end{align}
which likewise does not vanish at complex infinity.

The most realistic model for spherical nuclei, and the one most widely used in the literature, is the two-parameter Fermi (2pF) distribution,
\begin{align}\label{2pF}
\rho_{\mathrm{2pF}}(r) =&\ \frac{\rho_0}{1+e^{(r-c)/a}}\,.
\end{align}
As demonstrated in Appendix~\ref{app:fermi}, this distribution also does
not have the correct analytic properties for the physical values of the model
parameters $c>0$ and $a>0$.

Out of numerous models of the nuclear-charge distribution used in the literature, only
the exponential (1pE) model
\begin{align}
\rho_{\mathrm{1pE}}(r) =&\ \frac{\lambda^3}{8\,\pi}\, e^{-\lambda\,r}
\end{align}
satisfies the analytic properties required for the physical form factors. Its
Fourier transform
\begin{align}
F_{\mathrm{1pE}}(k^2) =&\ \frac{\lambda^4}{(k^2+\lambda^2)^2}
\end{align}
vanishes at the complex infinity and has a second-order pole at $k^2 = -\lambda^2$.
The 1pE distribution was used in Refs.~\cite{pachucki:23:prl,pachucki:25:recfns} to calculate the nuclear recoil effect for nuclei of finite size. However, this model does not provide a realistic description of the charge distribution of actual nuclei, particularly heavy ones. An exception is the proton, whose form factor is very close to that of the exponential model.

In the present work, we introduce a novel model of the nuclear charge distribution, which we call the two-parameter Gamma (2p$\Gamma$) model. This model can reproduce the first two even moments of the nuclear charge distribution and thus provides an accurate description of finite nuclear-size effects on atomic energy levels. The Gamma model closely resembles the well-known 2pF distribution but has the correct analytic properties in the complex $k^2$ plane, allowing us to use it in calculations of the nuclear recoil effect for nuclei of finite size.

The structure of the paper is as follows.
We start with formulating the Gamma model for the nuclear charge density and the corresponding
electrostatic potential
in the coordinate space in Sec.~\ref{sec:2pG:1}.
In Sec.~\ref{sec:2pG:2}, we transform the charge density to momentum space and study the analytic properties of
the momentum-space form factor in the complex $k^2$ plane. Next, Sec.~\ref{sec:2pG:3} examines the photon propagator with the nuclear charge form factor at one vertex, which is a key ingredient in the treatment of nuclear recoil.
The theory of the nuclear recoil effect is summarized in Sec.~\ref{sec:rec}.
Section~\ref{sec:num} presents our numerical results and Sec.~\ref{sec:code} gives a short introduction to the Fortran code RECFNS.
The relativistic units ($\hbar=c=m=1$, where $m$ is the mass of the bound lepton) and the Heaviside charge units ($ \alpha = e^2/4\pi$)
are used throughout this paper, unless explicitly specified.

\section{Gamma model in coordinate space}
\label{sec:2pG:1}
We introduce the two-parameter Gamma (2p$\Gamma$) model for the description of the nuclear charge density, defined as
\begin{align}
\rho_{\mathrm{2p}\Gamma}(\kappa,R;r) =&\ \frac{3\,\kappa^2}{4\,\pi\,R^3\,(\kappa+1)\,(\kappa+2)}\,Q\left(\kappa,x\right)
\,,
\end{align}
where $\kappa$ and $R$ are the model parameters,
$x \equiv \kappa\, r/R$ and
$Q$ is the regularized upper incomplete Gamma function,
\begin{align}
Q(a,x) \equiv \frac{\Gamma(a,x)}{\Gamma(a)}\,.
\end{align}
For integer $\kappa=n+1$, the above expression simplifies since $Q$ reduces to
\begin{align}
Q(n+1,x) = e^{-x}\,\sum_{k = 0}^{n} \frac{x^k}{k!}\,.
\end{align}
The 2p$\Gamma$ charge density is positive and normalized as $\int d^3 r\,\rho(r) = 1$.

The shape of $\rho_{\mathrm{2p}\Gamma}$ is similar to that of the well-known 2pF distribution (\ref{2pF}).
This can be readily seen for integer $\kappa$: for large $\kappa$, the truncated exponential
series nearly cancels $e^{-x}$ inside the nucleus, resulting in an approximately flat central density that then falls exponentially at the surface.
In the limit of large $\kappa$, the 2p$\Gamma$ density coincides with the hard sphere model,
\begin{align}
\rho_{\mathrm{2p}\Gamma}(\infty,R;r) = \Theta(R-r)\,\frac{3}{4\,\pi\,R^3}\,.
\end{align}
In another limiting case, $\kappa = 1$, the 2p$\Gamma$ model becomes the well-known
exponential (1pE) distribution,
\begin{align}
\rho_{\mathrm{2p}\Gamma}(1,R;r) =
\frac{e^{-x}}{8\,\pi\,R^3}\,.
\end{align}
All nuclear charge moments of the Gamma model have a very simple form. The mean square radius is
\begin{align}\label{eq:rms}
\langle r^2\rangle =&\ \frac{3}{5}\,R^2\,\frac{(\kappa+3)\,(\kappa+4)}{\kappa^2}\,,
\end{align}
whereas for an arbitrary power
\begin{align}
\langle r^m \rangle =&\ \frac{3}{m+3}\,R^m \frac{\left( \kappa+3\right)_m}{\kappa^m}\, \,,
\end{align}
where $\left( a\right)_n = \Gamma(a+n)/\Gamma(a) = a(a+1)\ldots(a+n-1)$ is the Pochhammer symbol.

\begin{figure*}
\centerline{
\resizebox{0.9\textwidth}{!}{%
  \includegraphics{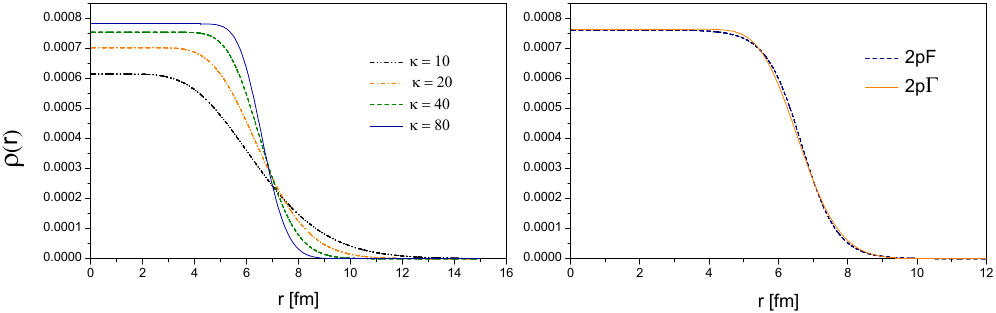}
}}
 \caption{
Left panel: $\rho_{\mathrm{2p}\Gamma}(\kappa,R;r)$
for $R = 6.65$~fm and $\kappa = 10$, 20, 40, and 80. Right panel: comparison
of the 2p$\Gamma$ and 2pF nuclear-charge densities
for $r_C = 5.5030$~fm and $\eta = 1.06315$; nuclear model parameters are
$R = 6.65659$~fm, $\kappa = 52$, and $c= 6.6652$~fm, $a= 0.51242$~fm.
\label{fig:density}}
\end{figure*}

The parameters of the Gamma model have an intuitive physical interpretation.
The parameter $R$ can be regarded as the nuclear ``radius'',
whereas $\kappa$ is a dimensionless parameter defining the surface diffuseness of the nuclear charge
distribution, see the left panel of Fig.~\ref{fig:density}.

The two-parameter 2p$\Gamma$ distribution can reproduce the first two even moments
of the nuclear-charge distribution. If $ \lbr r^2\rbr$ and $ \lbr r^4\rbr$
are known, the parameter $\kappa$ can be obtained
as
\begin{align}
\kappa = \frac{-147 \eta ^4+\sqrt{441 \eta ^8+7350 \eta ^4+625}+275}{42 \eta ^4-50}\,,
\end{align}
where
\begin{align}
\eta \equiv \frac{\lbr r^4\rbr^{1/4}}{ \lbr r^2\rbr^{1/2}}\,.
\end{align}
The parameter $R$ is then obtained from $ \lbr r^2\rbr$  and $\kappa$ according to Eq.~(\ref{eq:rms}). For realistic
nuclear-charge distributions,
$\eta$ typically varies within the range
from $(5/2)^{1/4} = 1.257\ldots$, which corresponds to the exponential
distribution ($\kappa = 1$), to $(50/42)^{1/4} = 1.044\ldots$, which corresponds to the hard sphere
distribution ($\kappa = \infty$).

The electrostatic potential induced by the 2p$\Gamma$ charge density is given by
\begin{align}
V_{\mathrm{2p}\Gamma}(\kappa,R;r)  =&\ -\frac{Z\,\alpha}{r}
\Big[1 -
Q(\kappa+3,x)
\nonumber \\ &
+ \frac{3\,x}{2(\kappa+2)}\,Q(\kappa+2,x)
\nonumber \\ &
- \frac{x^3}{2\kappa(\kappa+2)(\kappa+1)}\,Q(\kappa,x)
\Big]\,. \label{08}
\end{align}
The potential $V_{\mathrm{2p}\Gamma}$ is finite at the origin and
asymptotically Coulomb-like. For integer $\kappa$ it reduces to
\begin{align}
V_{\mathrm{2p}\Gamma}(\kappa,R;r)  =&\ -\frac{Z\,\alpha}{r}\,\big[1 - e^{-x}\,P_\kappa(x)\big]\,,
\end{align}
where $P_\kappa(x)$ is a polynomial of degree $\kappa$ with $P_\kappa(0)=1$.

In practical calculations 
with given $\lbr r^2\rbr$ and $\eta$ parameters,
it is often sufficient to use the nearest integer
value of $\kappa$ and readjust $R$ to yield
exactly the mean-square radius.
Even with an integer value of $\kappa$, the relative difference of
the fns corrections calculated with the 2pF and 2p$\Gamma$ models
is very small. In the following we will always assume that $\kappa$ is rounded
to the nearest integer, since this greatly simplifies computations in the
momentum space required in the treatment of
the nuclear recoil effect.

It is instructive to compare the 2p$\Gamma$ distribution with the commonly used
two-parameter Fermi (2pF) distribution. Such a comparison is presented in the right panel of
Fig.~\ref{fig:density}.
We observe that if both distributions correspond to the same values of
$\lbr r^2\rbr$ and $\eta$, the two densities are  hardly distinguishable visually,
even when $\kappa$ is rounded off to the nearest integer.

Table~\ref{tab:1} compares results obtained for three different nuclear models
for the ground state of the electronic and muonic H-like $^{208}\mathrm{Pb}$ ions.
For the purposes of the present calculations, the nuclear parameters of the
isotopes are assumed
to be given exactly by the values listed in the table.
The table presents numerical values for the fns correction
to the Dirac energy level ($E_{\rm fns}$) and the leading nuclear recoil correction
\begin{align}
E_\mathrm{L} = \Big< \frac{{\vec{p}}\,^2}{2M}\Big>\,, \label{18}
\end{align}
where $M$ is the nuclear mass. In contrast to the complete nuclear recoil
correction examined in Sec.~\ref{sec:rec}, the leading recoil correction $E_\mathrm{L}$
can be computed for an arbitrary nuclear charge density.
We observe that the results obtained with the 2pF and 2p$\Gamma$ models are very close
to each other, in contrast to those obtained with the 1pE model.


%
%
%
\begin{table}[t]
\caption{
Finite nuclear size correction $E_{\rm fns}$ and the leading recoil correction $E_\mathrm{L}$ given by Eq.~(\ref{18})
for the
$1s$ state of the electronic and muonic H-like $^{208}\mathrm{Pb}$ ions,
for different models of the nuclear charge distribution.
Nuclear parameters are:
$r_C = 5.5062~\mathrm{fm}$ \cite{sun:25}, $\eta = 1.0628$ \cite{yerokhin:26:rms}.
\label{tab:1}
}
\begin{center}
\begin{ruledtabular}
\begin{tabular}{llcc}
 Term
 &  Model
              &  \multicolumn{1}{c}{$e\,^{208}\mathrm{Pb}^{81+}$ [eV]}
                &  \multicolumn{1}{c}{$\mu\,^{208}\mathrm{Pb}^{81+}$ [keV]}
\\ \hline\\[-9pt]
$E_{\rm fns}$
&2pF         & 67.2842  & 10481.74 \\
&2p$\Gamma$  & 67.2824  & 10480.64 \\
&1pE         & 65.6943  &  9921.81 \\[3pt]
$E_{L}$
&2pF         &  0.566294 &  2.98695 \\
&2p$\Gamma$  &  0.566295 &  2.98821 \\
&1pE         &  0.567314 &  3.52464 \\
\end{tabular}
\end{ruledtabular}
\end{center}
\end{table}

\section{Gamma model in momentum space}
\label{sec:2pG:2}

We now examine the electrostatic nuclear form factor, which is obtained as the Fourier transform
of the nuclear charge density according to Eq.~(\ref{eqFF}).
When working in momentum space, it will be convenient to use the parameter $\lambda$ instead of $R$,
\begin{align}
\lambda \equiv \frac{\kappa}{R}\,,
\end{align}
so that the nuclear charge density becomes
\begin{align}
\rho_{{\rm 2p}\Gamma}(r) \equiv \rho_{\kappa,\lambda}(r) =&\
\frac{3\,\lambda^3}{4\,\pi}\,\frac{\Gamma(\kappa,\lambda\,r)}{\Gamma(\kappa+3)}\,.
\end{align}

We derive an expression for the nuclear form factor $F \equiv F_{\kappa,\lambda}$, with $y=k/\lambda$, as
\begin{align}
F_{\kappa,\lambda}(k^2) =&\
\int d^3r\,e^{i\,\vec k\cdot\vec r}\, \frac{3\,\lambda^3}{4\,\pi}\,\frac{\Gamma(\kappa,\lambda\,r)}{\Gamma(\kappa+3)}
\nonumber \\ =&\
\frac{3}{y^3\,(\kappa+2)\,2\,i}\,\bigg(\frac{1}{\kappa\,(1-i\,y)^\kappa}
\nonumber \\ &\
-\frac{1}{(\kappa+1)\,(1-i\,y)^{\kappa+1}}\biggr)
+ \mathrm{c.c.}
\nonumber \\ =&\
{}_2F_1\bigg(\frac{\kappa+3}{2}, \frac{\kappa+4}{2}, \frac{5}{2}; - y^2\bigg)
\,,
\end{align}
where the last equality can be verified by comparing the Taylor series of both sides of the equation.
So, the momentum-space form factor is
\begin{align}\label{eq13}
F_{\kappa,\lambda}(k^2) =&\ {}_2F_1\bigg(\frac{\kappa+3}{2}, \frac{\kappa+4}{2}, \frac{5}{2}; - \frac{k^2}{\lambda^2}\bigg)\,.
\end{align}
This form factor has the correct analytic properties expected for an arbitrary particle, including nuclei.
It is an analytic function in the complex $k^2$ plane, with the exception of
a branch cut along $k^2 \le -\lambda^2$ for noninteger $\kappa$,
which reduces to a pole of order $\kappa+1$ at $k^2=-\lambda^2$ for integer $\kappa$.
The form factor vanishes at large $k^2$ along any direction in the complex plane,
\begin{align}
F_{\kappa,\lambda}(k^2) \underset{k^2 \to \infty}{=}  &\ \frac{3\,\sin(\pi\,\kappa/2)}{\kappa\,(\kappa+2)}\,\biggl(\frac{\lambda}{k}\biggr)^{\kappa+3}\,.
\end{align}
For even integer $\kappa$ this leading term vanishes and the form factor falls off as $(\lambda/k)^{\kappa+4}$.
The small $k$-expansion of the form factor is
\begin{align}
F_{\kappa,\lambda}(k^2) =&\ 1 -\frac{(\kappa+3)\,(\kappa+4)}{10}\,\frac{k^2}{\lambda^2}
\nonumber \\ &\
+ \frac{(\kappa+3)\,(\kappa+4)\,(\kappa+5)\,(\kappa+6)}{280}\,\frac{k^4}{\lambda^4} + \ldots\,,
\end{align}
where the expansion coefficients are connected to $\langle r^{2n}\rangle$ in the standard way.

\subsection{Integer $\bm \kappa$}

We now address the case of an integer $\kappa$,
which will be extensively used in this work.
In the simplest case of $\kappa=1$, we have
\begin{align}
\rho_{1,\lambda}(r) =&\ \frac{\lambda^3}{8\,\pi}\,e^{-\lambda\,r}
\end{align}
and
\begin{align}
F_{1,\lambda}(k^2) = \frac{\lambda^4}{(\lambda^2+k^2)^2}.
\end{align}

Now we consider the case of an arbitrary integer $\kappa=n+1$.
The incomplete Gamma function $\Gamma(n+1,x)$ can be represented as
\begin{align}
\Gamma(n+1,x) =&\ (-1)^{n}\,x^{n+1}\frac{d^n}{d x^n}\, \frac{e^{-x}}{x}\,.
\end{align}
For $x=\lambda\,r$, it translates to
\begin{align}
\Gamma(n+1,\lambda\,r) =&\ (-1)^{n}\,\lambda^{n+1}\frac{d^n}{d \lambda^n}\, \frac{e^{-\lambda\,r}}{\lambda}\,.
\end{align}
Therefore, the charge density in coordinate space can be written as
\begin{align}\label{eq20}
\rho_{n+1,\lambda}(r) =&\
(-1)^{n}\,\frac{3\,\lambda^{n+4}}{4\,\pi\,\Gamma(n+4)}\,\frac{d^n}{d \lambda^n}\, \frac{e^{-\lambda\,r}}{\lambda}
\nonumber \\ =&\
\frac{(-1)^{n}}{(4)_n}\,\lambda^{n+4}\,\frac{d^n}{d \lambda^n}\, \frac{\rho_{1,\lambda}(r)}{\lambda^4}
\,.
\end{align}
The same relation holds for the form factor,
\begin{align}\label{eq21}
F_{n+1,\lambda}(k^2) =&\ \frac{(-1)^{n}}{(4)_n}\,\lambda^{n+4}\,\frac{d^n}{d \lambda^n}\, \frac{F_{1,\lambda}(k^2)}{\lambda^4}\,.
\end{align}

\section{Photon propagator}
\label{sec:2pG:3}

We now examine the photon propagator with a nuclear charge form factor $F$ at one vertex. This
propagator is needed to describe the nuclear recoil effect for an extended-size nucleus.
Because of relativistic covariance, any form factor for an arbitrary momentum exchange
is a function of  $k^2-\omega^2$, where $k^2 = \vec k^{\,2}$ and $\omega = k^0$.
With this in mind, the transverse part of the photon propagator in the Coulomb gauge,
with the nuclear charge form factor at one vertex can be derived  as \cite{pachucki:23:prl}
\begin{align}\label{eq22}
G^{ij}_C(\omega,\vec r) = \delta^{ij}\,{\cal D}(\omega, r) +
 \frac{\nabla^i\nabla^j}{\omega^2}\big[ {\cal D}(\omega, r) - {\cal D}(0, r)\big]\,,
\end{align}
where
\begin{align}
{\cal D}(\omega, r)= \int\frac{d^3k}{(2\,\pi)^3}\,e^{i\,\vec k\cdot\vec r}\;\frac{F(k^2-\omega^2)}{\omega^2-k^2}\,.
\end{align}
We note that ${\cal D}(0, r)$ is connected to the
nuclear potential given by Eq.~(\ref{08}) by
\begin{align}
{\cal D}(0, r) = \frac1{4\,\pi\,Z\,\alpha}\, V_{\mathrm{2p}\Gamma}(\kappa,R;r)\,.
\end{align}

Using the identity $\nabla^j\,f(r) = \frac{r^j}{r}\,f^{\prime}(r) \equiv \hat{r}^j\,f^{\prime}(r)$, we can rewrite Eq.~(\ref{eq22})
as
\begin{align}\label{eq24}
G^{ij}_C(\omega,\vec r) = \delta^{ij}\,{\cal D}(\omega, r) +
 \nabla^i \hat{r}^j\, {\cal F}(\omega, r)\,,
\end{align}
where
\begin{align}\label{eq25}
{\cal F}(\omega, r)=
\frac1{\omega^2}\Big[ {\cal D}^{\prime}(\omega, r) - {\cal D}^{\prime}(0, r)\Big]\,.
\end{align}
So, we expressed the photon propagator in terms of two scalar functions,
${\cal D}(\omega, r)$ and ${\cal F}(\omega, r)$.
For the point nucleus, these functions are given by simple formulas \cite{artemyev:95:pra}
\begin{align}
{\cal D}(\omega, r)&\ = -\frac{e^{i \omega\,r}}{4\pi r}\,, \\
{\cal F}(\omega, r)&\ = -\frac{1 - e^{i\omega\,r} (1- i\omega r )}{4 \pi \omega^2 r^2}\,.
\end{align}
Here and in what follows, $e^{i\omega r}$ should be understood as $e^{i\sqrt{\omega^2+i0}\,r}$.

For the Gamma model with parameters $\kappa$ and $\lambda$, we may use the general expression
for the form factor (\ref{eq13}) to obtain
\begin{align}
{\cal D}_{\kappa,\lambda}(\omega, r)=&\
-\frac{e^{i \omega\,r}}{4\pi r}\ + \frac{1}{2\,\pi^2\,r}\,\int_0^\infty dk\,\frac{k}{\omega^2-k^2}\,\sin(k\,r)\,
 \nonumber \\[2pt] & \times
\bigg[ {}_2 F_1\bigg(\frac{\kappa+3}{2}, \frac{\kappa+4}{2}, \frac{5}{2}; \frac{\omega^2-k^2}{\lambda^2}\bigg)-1\bigg]
\,.
\end{align}
The above expression can be used to compute ${\cal D}_{\kappa,\lambda}$ numerically, but this involves a numerical Fourier-transform integration over $k$.
This approach may be worth exploring, particularly in generalized cases where, for example, the electron vacuum polarization is included in the photon propagator.
In the present work, however, we restrict ourselves to the case of integer $\kappa$, for which recurrence relations can be derived that considerably simplify the numerical evaluation.

For integer $\kappa$, ${\cal D}_{\kappa,\lambda}$ can be expressed in terms of elementary functions, as it satisfies the same relations as the charge density, Eq.~(\ref{eq20}), and the form factor, Eq.~(\ref{eq21}).
Specifically,
\begin{align}\label{eq29}
{\cal D}_{n+1,\lambda}(\omega,r) =&\ \frac{(-1)^{n}}{(4)_n}\,\lambda^{n+4}\,\frac{d^n}{d \lambda^n}\,
   \frac{{\cal D}_{1,\lambda}(\omega,r)}{\lambda^4}\,.
\end{align}
The expression for ${\cal D}_{1,\lambda}$ was obtained in our previous work \cite{pachucki:23:prl}. For our
present purposes we write it as
\begin{align}
{\cal D}_{1,\lambda}(\omega,r) =&\
 \frac{\lambda^3}{8\pi r}\,
   \frac{d}{d\lambda}
      \frac{e^{i\omega r} - e^{i\sqrt{\omega^2-\lambda^2}\,r}}{\lambda^2}\,.
\end{align}
Inserting this expression into Eq.~(\ref{eq29}) and rearranging terms, we arrive at
\begin{align}
{\cal D}_{n+1,\lambda}(\omega,r) =&\ -\frac{e^{i\omega r}}{4\pi r}
+  \frac{\lambda^4}{8\pi r} \frac{\lambda^n}{(4)_n}
 \nonumber \\ &
\hspace{-15ex}  \times
 \bigg\{
 \bigg(\!-\frac{d}{d\lambda}\biggr)^{n+1} \frac{e^{i\sqrt{\omega^2-\lambda^2}\,r}}{\lambda^3}
-  \bigg(\!-\frac{d}{d\lambda}\biggr)^{n}\frac{e^{i\sqrt{\omega^2-\lambda^2}\,r}}{\lambda^4}
 \bigg\}\,.
\end{align}
Let us introduce two functions, $\tilde f$ and  $\tilde g$
\begin{align}\label{eq44}
\tilde f(n,m) \equiv&\  \frac{\lambda^{n+m}}{\Gamma(n+1)}\,\left(-\frac{d}{d\lambda}\right)^n \frac{e^{i\sqrt{\omega^2-\lambda^2}\,r}}{\lambda^m} \,, \\
\label{eq44g}\tilde g(n,m) \equiv&\  \frac{\lambda^{n+m+1}}{\Gamma(n+1)}\,\left(-\frac{d}{d\lambda}\right)^n \frac{e^{i\sqrt{\omega^2-\lambda^2}\,r}}{\lambda^m}\,\frac{i}{\sqrt{\omega^2-\lambda^2}} \,,
\end{align}
then ${\cal D}$ and ${\cal D}'$ can be expressed as
\begin{align}\label{eq45}
{\cal D}_{n+1,\lambda}(\omega,r)  =&
 -\frac1{4\pi\,r}\,
 \bigg\{ e^{i\omega r}
\nonumber \\ &
+ \frac{3\,\big[ \tilde f(n,4) - (n+1)\,\tilde f(n+1,3) \big]}{(n+3)\,(n+2)\,(n+1)}\bigg\}\,,\\
\label{eq45p}{\cal D}'_{n+1,\lambda}(\omega,r) =&\ -\frac{1}{r}\,{\cal D}_{n+1,\lambda}(\omega,r)  -\frac1{4\pi\,r}\, \bigg\{ i\,\omega\,e^{i\omega r}
\nonumber \\ &\
+ \frac{\omega^2}{\lambda}\frac{3\,\big[ \tilde g(n,4) - (n+1)\,\tilde g(n+1,3) \big]}{(n+3)\,(n+2)\,(n+1)}
\nonumber \\ &\
-\lambda\,\frac{3\,\big[ \tilde g(n,2) - (n+1)\,\tilde g(n+1,1) \big]}{(n+3)\,(n+2)\,(n+1)}
 \bigg\} \,.
\end{align}
Computations of the functions $\tilde f(n,m)$ and $\tilde g(n,m)$ are conveniently carried out
by recurrence relations derived in Appendix~\ref{app:1}.

\section{Nuclear recoil}
\label{sec:rec}

The nuclear recoil correction to hydrogenic energy levels can be represented in the following compact form \cite{pachucki:24:hfs}
 \begin{align}\label{eq46}
E_\mathrm{rec} =&\
\frac{i}{2\,\pi M} \int_s d\omega\,  \bra{a} D_T^j(\omega) \,G(\vare_a + \omega)\,D_T^j(\omega) \ket{a}\,,
\end{align}
where
$M$ is the nuclear mass, $\vare_a$ is the Dirac energy of the state $a$,
$G(E) = [E-H_D]^{-1}$ is the Dirac-Coulomb Green function,
\begin{align}
D_T^j(\omega,\vec r) = -4\pi Z\alpha \, \alpha^i \, G_{T}^{ij}(\omega,\vec{r}),
\end{align}
and
$G_T$ is the photon propagator with the nuclear charge form factor $F$ in the temporal gauge
\begin{align}
G_T^{ij}(\omega, \vec r) =&\ \int\frac{d^3k}{(2\,\pi)^3}\,e^{i\,\vec k\cdot\vec r}\,\frac{F(k^2-\omega^2)}{\omega^2-k^2}\,\biggl(\delta^{ij}-\frac{k^i\,k^j}{\omega^2}\biggr)\,.
\end{align}
The subscript $s$ in Eq.~(\ref{eq46}) denotes the Feynman contour with symmetric integration around the pole at $\omega=0$.

The expression (\ref{eq46}) can be rewritten in the Coulomb gauge as \cite{
shabaev:85,shabaev:88,pachucki:95,shabaev:98:rectheo,pachucki:23:prl,pachucki:24:hfs}
\begin{align}\label{eq49}
E_{\rm rec} =&\, \frac{i}{2\pi M}\, \int_{-\infty}^{\infty}d\omega\,
\sum_n
 \frac1{\vare_a + \omega - \vare_n(1-i0)}
\nonumber \\ & \times
 \bra{a} \vec{p} - \vec{D}_C(\omega) \ket{n}
   \bra{n} \vec{p} - \vec{D}_C(\omega) \ket{a}\,,
\end{align}
where
$\vec{p}$ is the momentum operator,
$\vec{D}_C(\omega)$ is
connected with the transverse part of the photon propagator in the Coulomb gauge $G_C^{ij}$ by
\begin{equation}
D_C^j(\omega) = -4\pi Z\alpha \, \alpha^i \, G_C^{ij}(\omega,\vec{r})\,,
\end{equation}
and $G_C^{ij}$ is defined by Eq.~(\ref{eq22}).

For numerical calculations, it is convenient to separate the nuclear recoil correction (\ref{eq49})
into several parts, which are induced by the exchange of Coulomb photons only,
by Coulomb photons and one transverse photon ($E_{\rm tr1}$),
and by Coulomb photons and two transverse photons ($E_{\rm tr2}$).
Furthermore, the Coulomb-photon part is separated into the leading-order part $E_{\rm L}$
and the higher-order Coulomb-photon part $E_{\rm C}$. We thus write
\begin{align}
E_{\rm rec} = E_{\rm L} + E_{\rm C} + E_{\rm tr1} + E_{\rm tr2}\,.
\end{align}
The leading-order contribution (the same as in Eq.~(\ref{18})) is
\begin{align}
E_{\rm L}  = \,\bra{a}  \frac{\vec{p}\,^2 }{2M}\ket{a}\,.
\end{align}
The remaining Coulomb-photon contribution is given by
\begin{align}
E_{\rm C}  = -\frac{1}{M}\, \sum_{\vare_n < 0} \bra{a} \vec{p} \ket{n}\, \bra{n} \vec{p} \ket{a}\,,
\end{align}
where the summation is performed over the {\em negative-energy continuum} part of the Dirac spectrum.
The one-transverse-photon and the two-transverse-photon contributions are given by
\begin{align}
E_{\rm tr1} = &\,
     -\frac{i}{\pi M}\, \int_{-\infty}^{\infty}d\omega\,
\sum_n
 \frac1{\vare_a + \omega - \vare_n(1-i0)}\,
\nonumber \\ & \times
 \bra{a} \vec{p}\ket{n}\,
   \bra{n}\vec{D}_C(\omega) \ket{a}\,, \\
E_{\rm tr2} = &\,
     \frac{i}{2\pi M}\, \int_{-\infty}^{\infty}d\omega\,
\sum_n
 \frac1{\vare_a + \omega - \vare_n(1-i0)}\,
\nonumber \\ & \times
  \bra{a} \vec{D}_C(\omega)\ket{n} \,
   \bra{n} \vec{D}_C(\omega) \ket{a}\,.
\end{align}

For numerical calculations, it is advantageous to rewrite matrix elements of
$\vec{p}$ and $\vec{D}_C$ using commutator relations with the Dirac Hamiltonian $H_D =
\vec{\alpha}\cdot\vec{p} + \beta m + V_{\rm nuc}(r)$.
Specifically, one has \cite{shabaev:94:rec}
\begin{align}
\bra{a}\vec{p}\ket{n} = &\ \bra{a}\Big[
\frac12\, \left\{ \balpha, H_D \right\} - \balpha\, V_{\rm nuc}(r)\Big]\ket{n}
 \nonumber \\
 =&\  \bra{a} \balpha\,\Big[ \frac{\vare_a+\vare_n}{2} -V_{\rm nuc}(r)\Big] \ket{n}
 \,,
\end{align}
where $\vare_a$ and $\vare_n$ are the Dirac energies of the states $a$ and $n$, respectively.
Similarly, for the matrix elements of $\vec{D}_C$ one obtains \cite{artemyev:95:pra}, see Eqs.~(\ref{eq22})-(\ref{eq25})
\begin{align}
\bra{a}\vec{D}_C\ket{n} =
 -4\pi Z\alpha\, \bra{a} \Big[ \balpha\, {\cal D}(\omega,r)
  + \big( \balpha\cdot\bNabla\big) \hat{\vec{r}}\, {\cal F}(\omega,r) \Big] \ket{n}\,.
\end{align}
Using the fact that $\big(\balpha\cdot\bNabla\big) f(r) = i[H_D,f(r)]$, one arrives at
\begin{align}
\bra{a}\vec{D}_C\ket{n} =
 -4\pi Z\alpha\, \bra{a} &\  \Big[ \balpha\, {\cal D}(\omega,r)
 \nonumber \\ &
  + i \big(\vare_a-\vare_n\big) \hat{\vec{r}}\, {\cal F}(\omega,r) \Big] \ket{n}\,.
\end{align}
Further reduction of matrix elements to radial integrals is carried out by using the
standard Racah algebra, with explicit expressions presented in
Ref.~\cite{yerokhin:23:recoil}.

In computations of the one-transverse-photon and the two-transverse-photon contributions,
one needs to evaluate integrals over $\omega$ numerically. In order to facilitate
the computations, it is advantageous to make a Wick rotation of the integration
contour in the $\omega$ complex plane, which leads to the appearance of pole contributions
from states bound as deeply as, or more deeply than, the reference state $a$.
The general formula of the Wick rotation for a typical integral contribution is
\begin{align}
\frac{i}{2\pi} \int_{-\infty}^{\infty} & d\omega \sum_n
\frac{f(\omega)}{\vare_a + \omega - \vare_n(1-i0)}
=
\sum_{0 < \vare_n \leq \vare_a} a_n\,
 f(\Delta_{na})
\nonumber \\    &
     -\frac{1}{\pi}\, \int_{0}^{\infty}d\omega\,
\sum_{n} \frac{\Delta_{an}}{\Delta_{an}^2+\omega^2}
   \, f(i\omega) \,,
\end{align}
where $\Delta_{an} = \vare_a-\vare_n$, it is assumed that $f(i\omega) = f(-i\omega)$ is real, and
$a_n = 1$ for $\vare_n \neq \vare_a$ and $a_n = \nicefrac12$ for $\vare_n = \vare_a$.

For numerical calculations of the recoil corrections one needs to sum over the spectrum of the
Dirac equation. To this end, we use the dual kinetic balance (DKB) basis-set  method
\cite{shabaev:04:DKB} with basis functions constructed with $B$-splines.
In order to ensure the numerical stability of the method in the region of low
nuclear charges, the DKB method was implemented in the quadruple-precision arithmetic
(about. 32 decimal digits), following Ref.~\cite{yerokhin:15:recprl}.

\section{Numerical results}
\label{sec:num}

Numerical results for the nuclear recoil correction to the energy level $E_{\rm rec}$, are conveniently represented
in terms of the dimensionless function $P_{\rm rec}$ defined as
\begin{align}\label{eq50}
E_{\rm rec} = m c^2\,\frac{m}{M}\,\frac{(\Za)^2}{2n^2}\,P_{\rm rec}(\Za)\,.
\end{align}
Here, $m$ is the mass of the corresponding lepton (muon for the muonic atoms and
electron for the electronic atoms), and $n$ is the principal quantum number of the reference state.
The function $P_{\rm rec}$ is defined so that it is unity in the nonrelativistic limit
(corresponding to the first-order in $m/M$ part of the reduced-mass prefactor).

Our numerical results obtained for the $n \leq 3$ states of the electronic and muonic H-like lead
ions are presented in Table~\ref{tab:2}.
We compare results obtained for three different nuclear models:
the point (pnt) nucleus, the one-parameter exponential (1pE) model, and the two-parameter
Gamma (2p$\Gamma$) model. The point-nucleus results are the same for the electronic and muonic atoms;
they are in agreement with the 1995 calculation by Artemyev et al.~\cite{artemyev:95:pra}.
The values obtained with the exponential model are in agreement with the results previously reported in
Ref.~\cite{yerokhin:23:recoil}.

Analyzing the results presented in Table~\ref{tab:2},
we observe that the values of $P_{\rm rec}$ deviate significantly from unity for the low-lying states of Pb, particularly the $s$ states. This indicates that relativistic effects are strong and that the nonrelativistic treatment is inadequate for describing the nuclear recoil effect in low-lying states of high-$Z$ ions. By contrast, relativistic effects decrease rapidly with increasing $l$  and the corresponding $P_{\rm rec}$ values approach unity.

Next, we find that the fns effect strongly affects the nuclear recoil correction in muonic atoms. It is particularly pronounced for the $1s$ state, reducing the point-nucleus result for $^{208}$Pb by a factor of six. Such a large effect is remarkable but not very surprising, given that the muon $1s$ wave function in Pb is localized predominantly within the nucleus. Nuclear-size effects also remain substantial for higher $ns$ states but gradually diminish with increasing $l$.

It is instructive to compare the results in Table~\ref{tab:2} obtained with the 2p$\Gamma$ and 1pE models. For the electronic Pb ion, the nuclear-model dependence is quite small, at the level of 0.1\% for the $1s$ state and even smaller for excited states. This indicates that the fns recoil contribution depends predominantly on the mean-square radius $\lbr r^2\rbr$, with higher nuclear moments inducing only a very small contribution. The situation is dramatically different for muonic Pb, where the difference between the 2p$\Gamma$ and 1pE results for the $1s$ state reaches 17\%. This indicates that higher nuclear moments play a major role in the fns recoil effect.

The difference between the 2p$\Gamma$ and 1pE results decreases in absolute magnitude for higher excited states. It is interesting to note, however, that the {\em relative} difference between the fns recoil contributions obtained with the two models 
{\em increases}. 
For example, for the $3d_{5/2}$ state in $\mu\,^{208}\mathrm{Pb}$, the fns recoil contributions obtained with the two models differ by more than a factor of two. This reflects the increasing sensitivity of higher-$l$ states to higher moments of the nuclear charge distribution.

%
%
%
\begin{table}[t]
\caption{
Nuclear recoil correction for
the electronic and muonic H-like $^{208}\mathrm{Pb}$ ions,
for different models of the nuclear charge distribution
(``pnt'' denotes the point nucleus).
Results are given in terms of the dimensionless function $P_{\rm rec}(\Za)$ defined in Eq. (\ref{eq50}).
Nuclear parameters are the same as in Table~\ref{tab:1}.
\label{tab:2}
}
\begin{center}
\begin{ruledtabular}
\begin{tabular}{llw{2.6}w{2.6}}
 State
 &  Model
              &  \multicolumn{1}{c}{$e\,^{208}\mathrm{Pb}^{81+}$ }
                &  \multicolumn{1}{c}{$\mu\,^{208}\mathrm{Pb}^{81+}$ }
\\ \hline\\[-9pt]
%
%
%
$1s$ & pnt & 1.5182  & 1.5182  \\
    & 1pE & 1.47090 & 0.29948 \\
    & 2p$\Gamma$ & 1.46941 & 0.25691 \\[2pt]
$2s$ & pnt & 1.5245  & 1.5245  \\
    & 1pE & 1.49093 & 0.58203 \\
    & 2p$\Gamma$ & 1.48987 & 0.52474 \\[2pt]
$3s$ & pnt & 1.38353 & 1.38353 \\
    & 1pE & 1.36081 & 0.70858 \\
    & 2p$\Gamma$ & 1.36009 & 0.65931 \\[2pt]
$2p_{1/2}$ & pnt & 1.17919 & 1.17919 \\
    & 1pE & 1.17635 & 0.82872 \\
    & 2p$\Gamma$ & 1.17626 & 0.83086 \\[2pt]
$3p_{1/2}$ & pnt & 1.15354 & 1.15354 \\
    & 1pE & 1.15128 & 0.89303 \\
    & 2p$\Gamma$ & 1.15121 & 0.88658 \\[2pt]
$2p_{3/2}$ & pnt & 1.01186 & 1.01186 \\
    & 1pE & 1.01184 & 0.83339 \\
    & 2p$\Gamma$ & 1.01184 & 0.84264 \\[2pt]
$3p_{3/2}$ & pnt & 1.03280 & 1.03280 \\
    & 1pE & 1.03279 & 0.89556 \\
    & 2p$\Gamma$ & 1.03279 & 0.89575 \\[2pt]
$3d_{3/2}$ & pnt & 1.02194 & 1.02194 \\
    & 1pE & 1.02194 & 1.00623 \\
    & 2p$\Gamma$ & 1.02194 & 1.01275 \\[2pt]
$3d_{5/2}$ & pnt & 1.00078 & 1.00078 \\
    & 1pE & 1.00078 & 0.99083 \\
    & 2p$\Gamma$ & 1.00078 & 0.99637 \\
\end{tabular}
\end{ruledtabular}
\end{center}
\end{table}

Table~\ref{tab:3} compares our results obtained with the 2p$\Gamma$ model for $\mu^{35}$Cl and $\mu^{90}$Zr atoms with those of previous investigations~\cite{rathi:26,beyer:26}. In those works, the nuclear charge distribution was modeled using a simplified approach: the leading recoil contribution, $E_\mathrm{L}$, was calculated with the 2pF model, while the higher-order recoil corrections were evaluated using the 1pE model, following Refs.~\cite{pachucki:23:prl,yerokhin:23:recoil}. We observe good agreement between different calculations, although the model dependence turns out to have been slightly underestimated in the previous works.

%
%
%
\begin{table}[t]
\caption{
Nuclear recoil correction
for $\mu^{35}$Cl and $\mu^{90}$Zr atoms, in keV. Nuclear
parameters are the same as in Refs.~\cite{rathi:26,beyer:26}.
\label{tab:3}
}
\begin{center}
\begin{ruledtabular}
\begin{tabular}{lw{2.6}w{2.6}w{2.6}w{2.6}}
 State
              &  \multicolumn{2}{c}{$\mu^{35}$Cl }
                &  \multicolumn{2}{c}{$\mu^{90}$Zr}
\\ \cline{2-3} \cline{4-5}
\\[-9pt]
              &  \multicolumn{1}{c}{This work}
                &  \multicolumn{1}{c}{Ref.~\cite{rathi:26}}
              &  \multicolumn{1}{c}{This work}
                &  \multicolumn{1}{c}{Ref.~\cite{beyer:26}}
\\ \hline\\[-9pt]
$1s$        &       2.33270 &   2.330  &  3.30931 & 3.287\,(9)  \\
$2s      $  &       0.62257 &          &  1.10223 & 1.097\,(2)  \\
$2p_{1/2}$  &       0.66164 &   0.661  &  1.42118 & 1.421\,(1)  \\
$2p_{3/2}$  &       0.65925 &   0.659  &  1.40376 & 1.4037\,(5)  \\
$3p_{1/2}$  &       0.29398 &   0.294  &  0.63318 & 0.6329\,(2)  \\
$3p_{3/2}$  &       0.29328 &   0.293  &  0.62856 & 0.6285\,(2)  \\
$3d_{3/2}$  &       0.29335 &          &  0.63404 & 0.6339\,(1)  \\
$3d_{5/2}$  &       0.29310 &          &  0.63105 & 0.6309       \\
$4p_{1/2}$  &       0.16529 &   0.165  &  0.35633 \\
$4p_{3/2}$  &       0.16500 &   0.165  &  0.35446 \\
\end{tabular}
\end{ruledtabular}
\end{center}
\end{table}

\section{RECFNS code: minimal how-to guide}
\label{sec:code}

The numerical results presented in the previous section were obtained using the Fortran code RECFNS, which is available in the Supplementary Material. We now provide a brief guide that allows the user to run the code, reproduce the results reported above, and obtain additional data. More detailed information about the code is available in the README file, which is also included in the Supplementary Material.

The input of the code consists of several lines, the order of which is arbitrary,
with each line containing a symbolic key indicating which variable is specified in that line.
The execution time is several hours on a single processor, depending on the input parameters.
The usage of the code is best illustrated by examples.

As a first example, we calculate the $P_{\rm rec}$ value for the $1s$ state of
$e\,^{208}\mathrm{Pb}^{81+}$, given in Table~\ref{tab:2}. This is done as follows:
{\small
\begin{verbatim}
./recfns <<EOF
Z=82
State=1s1
RMS=5.5062
ETA=1.0628
EOF
\end{verbatim}
}
\noindent
Here, the first line specifies the nuclear charge $Z$, whereas the second line
contains the reference state (coded by three symbols: the principal
quantum number, the orbital angular momentum $l = 0,1,2,3$ in the standard
notation $s$, $p$, $d$, $f$, and twice the total angular momentum, $2j$).
The third and the fourth lines specify the values of the rms radius $r_C$
and the parameter $\eta$. With this input the code produces the 2p$\Gamma$ value in
Table~\ref{tab:2}.

In order to compute the analogous value for
$\mu\,^{208}\mathrm{Pb}$ listed in Table~\ref{tab:2},
we need to add an additional input line
that specifies the type of the atom:
\verb|ATOM=MU|.

Instead of specifying the parameter $\eta$, one can input explicitly the
parameter $\kappa$ of the 2p$\Gamma$ model. In particular, by setting
$\kappa = 1$, one obtains the one-parameter exponential 1pE model. So,
the following input
{\small
\begin{verbatim}
./recfns <<EOF
Z=82
State=2p3
RMS=5.5062
KAP=1
ATOM=MU
EOF
\end{verbatim}
}
\noindent
produces the 1pE value for the $2p_{3/2}$ state of
$\mu\,^{208}\mathrm{Pb}$ from Table~\ref{tab:2}.

The code can also compute directly corrections to energy in eV or keV units,
instead of the function $P_{\rm rec}$.
In order to do this, one needs to specify the units (eV or keV) and
provide the nucleus-to-electron mass ratio $M/m_e$. For example,
the following input
{\small
\begin{verbatim}
./recfns <<EOF
Z=82
State=1s1
RMS=5.5062
ETA=1.0628
ATOM=MU
DMratio=379037
UNITS=keV
EOF
\end{verbatim}
}
\noindent
produces the result $E_{\rm rec} = 2.65105$~keV.

\section{Summary}

We have introduced a novel model of the nuclear charge distribution, which we call the two-parameter Gamma (2p$\Gamma$) model.
In many respects, it provides an alternative to the widely used two-parameter Fermi (2pF) distribution.
Both models have similar shapes, but the Gamma model offers several advantages.
Among them is that all moments of the charge distribution, $\lbr r^n\rbr$, are given by simple analytical formulas,
and the form factor has a closed-form expression in momentum space.
With a negligible loss of accuracy, the parameter $\kappa$ of the Gamma model can be restricted to integer values,
in which case the charge density and the corresponding potential
take a simple form of an exponential multiplied by a polynomial,
and the form factor becomes a rational function of $k^2$ in momentum space.

The most important feature of the Gamma model is that its momentum-space form factor $F(k^2)$ has the correct analytic properties in the complex $k^2$ plane required for a physical charged particle. Specifically, it is analytic throughout the complex $k^2$ plane except for a branch cut along $k^2 \le -\lambda^2$, which reduces to a pole of order $\kappa+1$ at $k^2=-\lambda^2$ for integer $\kappa$, and it vanishes at complex infinity. These properties make the Gamma model suitable for calculating nuclear recoil effects in atomic systems within the framework of QED.

Using the Gamma model, we performed numerical calculations of the nuclear recoil effect for a realistic nuclear charge distribution and investigated the dependence of the theoretical predictions on the nuclear model. We found that nuclear-size recoil corrections in heavy muonic atoms are strongly sensitive to higher moments of the nuclear charge distribution, $\lbr r^n\rbr$ with $n>2$. This highlights the need to use a realistic nuclear model in theoretical calculations.

As recently demonstrated in Ref.~\cite{beyer:26}, an accurate treatment of nuclear recoil is essential for determining nuclear charge radii from muonic-atom spectra. Given the strong sensitivity of this effect to nuclear shape, such determinations require extensive calculations of the corresponding correction for different sets of nuclear parameters. To facilitate these calculations, we provide the Fortran code RECFNS in the Supplementary Material.

The calculation of the nuclear recoil contribution to the Lamb shift to first order in the electron-to-nucleus mass ratio
presented in this work can be extended to higher orders in the mass ratio, to the hyperfine splitting, and to radiative corrections, including the lepton self-energy and vacuum polarization,
using the formalism of Heavy-Particle QED \cite{pachucki:24:hpqed}. In the future, these extensions should enable highly accurate theoretical predictions for hydrogen-like systems with finite nuclear mass and size. The ultimate limitations on theoretical accuracy are then set by inelastic nuclear effects, such as nuclear polarizability.

\appendix

\section{Fermi distribution in momentum space}
\label{app:fermi}

The two-parameter Fermi distribution is defined in coordinate space as
\begin{align}
\rho_{\mathrm{2pF}}(r) =&\ \frac{\rho_0}{1+e^{(r-c)/a}}\,,
\end{align}
where
\begin{align}
\rho_0^{-1} =&\ -8\,\pi\,a^3\,\mathrm{Li}_3(-e^{c/a})\,.
\end{align}
The corresponding form factor in momentum space was derived in Ref.~\cite{maximon:66},
with the result
\begin{align}
F_{\mathrm{2pF}}(k^2) =&\ -\frac{4\,\pi\,\rho_0\,a^3}{b\,\beta} \Re\frac{\partial}{\partial\beta}\,\bigg[(1-i\,\beta)^{-1}
\nonumber \\ &\times {}_2F_1(1,1-i\,\beta, 2-i\,\beta; -b^{-1})\bigg]\,,
\end{align}
where $\beta = k\,a$ and $b=e^{-c/a}$, and the real part is taken assuming $\beta$ to be real.
The power series in $b^{-1}$ of the above expression
is convergent for $b > 1$  and the form factor vanishes for large $k^2$ in any direction
in the complex $k^2$ plane.
However, for the physical (real and positive) values of $a$ and $c$,
the parameter $b<1$. In this case, we can use a hypergeometric-function identity to obtain
\begin{align}
F_{\mathrm{2pF}}(k^2) =& -\frac{4\,\pi\,\rho_0\,a^3}{\beta} \Re\frac{\partial}{\partial\beta}\bigg[
\frac{\pi\, b^{-i\,\beta}}{i\,\sinh(\pi\,\beta)}
\nonumber \\ &
-\frac{1}{i\,\beta}\, {}_2F_1(1,i\,\beta, 1+i\,\beta; -b)\biggr]\,.
\end{align}
The first term in the brackets does not vanish for large complex values of $k^2$, while the second one does;
therefore, $F_{\mathrm{2pF}}$ does not have the analytic properties required for the form factor of a physical particle.

\section{Recurrence relations}
\label{app:1}

Here we develop the recurrence relations that allow efficient computation of the
functions ${\cal D}$ and ${\cal D}'$ according to Eqs.~(\ref{eq45}) and (\ref{eq45p}).
It will be convenient to first obtain recurrence relations for functions
$f(n,m)$ and $g(n,m)$ defined as
\begin{align}
f(n,m) =&\  \bigg(\! -\frac{d}{d\,\lambda}\bigg)^n\,  \frac{e^{-\wpp\,r}}{\lambda^m}\,,\\
g(n,m) =&\  \bigg(\! -\frac{d}{d\,\lambda}\bigg)^n\,  \frac{e^{-\wpp\,r}}{\lambda^m}\,\frac{1}{\wpp}\,,
\end{align}
where $\wpp = -i\sqrt{\omega^2-\lambda^2}$.
We note that $\wpp$ becomes real for purely imaginary $\omega$ and the
entire computation can then be performed using real arithmetic.

We start with deriving the recursion relations for $f(n) \equiv f(n,0)$ and $g(n) \equiv g(n,0)$.
We have
\begin{align}
f(1) =&\ \lambda\,r\,g(0)\,, \\
\wpp^2 \,g(1) =&\ \lambda\,g(0) + \lambda\,r\,f(0)\,.
\end{align}
Therefore,
\begin{align}
\bigg(\! -\frac{d}{d\,\lambda}\bigg)^{n-1}\!f(1) =&\ \bigg(\! -\frac{d}{d\,\lambda}\bigg)^{n-1}\!\lambda\,r\,g(0)\,, \\
\bigg(\! -\frac{d}{d\,\lambda}\bigg)^{n-1}\! \wpp^2 \,g(1) =&\ \bigg(\! -\frac{d}{d\,\lambda}\bigg)^{n-1}\!\lambda\,[g(0) + r\,f(0)]\,.
\end{align}
We thus obtain
\begin{align}
f(n) =&\ \lambda\,r\,g(n-1) - (n-1)\,r\,g(n-2)\,,
\end{align}
and
\begin{align}
\wpp^2 g(n)=&\,\lambda\,[g(n-1)+r\,f(n-1)]
\nonumber \\  &
-(n-1)\,[g(n-2)+r\,f(n-2)]
\nonumber \\  &
+(n-1)\,[2\,\lambda\,g(n-1)-(n-2)\,g(n-2)]\,.
\end{align}
We now obtain recursion relations for $f(n,m)$ and $g(n,m)$.
We observe that
\begin{align}
\lambda\,f(0,m) =&\ f(0,m-1)\,,\\
\bigg(\! -\frac{d}{d\,\lambda}\bigg)^n\, \lambda\,f(0,m) =&\ \bigg(\! -\frac{d}{d\,\lambda}\bigg)^n\, f(0,m-1)\,,
\end{align}
and the same for $g(n,m)$.
Therefore, we obtain
\begin{align}
\lambda\,f(n,m) =&\ f(n,m-1) + n\,f(n-1,m)\,, \\
\lambda\,g(n,m) =&\ g(n,m-1) + n\,g(n-1,m)\,.
\end{align}
Now we rewrite the obtained relations in terms of $\tilde f$ and $\tilde g$, defined by Eqs.~(\ref{eq44}) and (\ref{eq44g}).
We obtain
\begin{align}\label{a13}
\tilde f(n) =&\ \frac{\lambda\,r}{n}\,\big[ \tilde g(n-1) - \tilde g(n-2)\big]\,,\\
\tilde g(n)=&\ \frac{\lambda^2}{\wpp^2\,n}\,\big\{(2\,n-1)\,\tilde g(n-1) -(n-1)\,\tilde g(n-2)
\nonumber \\ &
+\lambda\,r\,[\tilde f(n-1) - \tilde f(n-2)]\big\}\,,
\label{a14}
\end{align}
and
\begin{align}
\label{a15}
\tilde f(n,m) =&\ \tilde f(n,m-1) + \tilde f(n-1,m),\\
\tilde g(n,m) =&\ \tilde g(n,m-1) + \tilde g(n-1,m)\,.
\label{a16}
\end{align}

The recurrence relations derived above allow us to efficiently compute the functions ${\cal D}$ and ${\cal D}'$ using Eqs.~(\ref{eq45}) and (\ref{eq45p}).
In our calculations, we start from the explicit expressions for $\tilde f(0,0)$, $\tilde f(1,0)$, $\tilde g(0,0)$, and $\tilde g(1,0)$, and use Eqs.~(\ref{a13}) and (\ref{a14}) to generate $\tilde f(n,0)$ and $\tilde g(n,0)$. We then combine these results with the explicit expressions
\begin{align}
\tilde f(0,m) =&\  e^{-\wpp\,r}\,,\\
\tilde g(0,m) =&\ e^{-\wpp\,r}\,\frac{\lambda}{\wpp}\,,
\end{align}
and apply Eqs.~(\ref{a15}) and (\ref{a16}) to compute the required values of $\tilde f(n,m)$ and $\tilde g(n,m)$.

\section{Small $\bm{\omega}$ expansion}
\label{app:2}

Here we derive the small-$\omega$ expansions of ${\cal D}$ and ${\cal F}$. These expansions are useful for efficient numerical calculations because the explicit expression for ${\cal F}$ in Eq.~(\ref{eq25}) contains an apparent singularity as $\omega \to 0$. The expansions show that ${\cal F}$ has a well-defined limit at $\omega = 0$.

We will obtain the expansions for purely imaginary frequencies, i.e., for ${\cal D}(i\omega,r)$ and ${\cal F}(i\omega,r)$ with real $\omega$,
since these are the quantities needed in actual computations.
First, we derive the expansion for ${\cal D}_{1,\lambda}$,
\begin{widetext}
\begin{align}
{\cal D}_{1,\lambda}(i\,\omega, r) =&\
 \frac{\lambda^3}{8\pi r}\,
   \frac{d}{d\lambda}
      \frac{e^{-\omega\, r} - e^{-\sqrt{\omega^2+\lambda^2}\,r}}{\lambda^2}
\nonumber \\ =&\
 -\frac1{4\pi}\, \left\{ \frac{e^{-\omega\, r}}{r}  + \lambda\,e^{-\lambda\,r}\left[
-\frac{1}{2} - \frac{1}{\lambda\,r} + \left(\frac{3}{4} + \frac{\lambda\,r}{4}\right)\,\frac{\omega^2}{\lambda^2}
- \left(\frac{5}{16} + \frac{5\,\lambda\,r}{16} + \frac{\lambda^2\,r^2}{16}\right)\,\frac{\omega^4}{\lambda^4} +  \ldots
\right]\right\}\,.
\end{align}
Next, we use the identity
\begin{align}
\bigg(\!-\frac{d}{d \lambda}\bigg)^n\, \frac{e^{-\lambda\,r}}{\lambda} =&\ \frac{\Gamma(n+1,\lambda\,r)}{ \lambda^{n+1}}\,
\end{align}
to derive the expansion for ${\cal D}_{n+1,\lambda}$
\begin{align}
{\cal D}_{n+1,\lambda}(i\,\omega,r)  =&\ \frac{\Gamma(4)}{\Gamma(n+4)}\,\lambda^{n+4}\,\bigg(\! -\frac{d}{d\,\lambda}\bigg)^n\, \frac{{\cal D}_{1,\lambda}(i\,\omega,r)}{\lambda^{4}}
=
-\frac1{4\pi\,r}\, \biggl\{ e^{-\omega\, r} +\sum_{i=0}^\infty (\omega\,r)^{2\,i}\,a_{2i} \biggr\}\,,
\end{align}
where the coefficients $a_{2i}$ are given by (with $x=\lambda\,r$)
\begin{align}
a_0 =&\   -\frac{x^3\,\Gamma(1 + n, x) - 3\,x\,\Gamma(3 + n, x) +  2\,\Gamma(4 + n, x)}{2\,\Gamma(4 + n)}\,,
\nonumber \\
a_2 =&\  -\frac{x^4\,\Gamma(1 + n, x) - 6\,x^2\,\Gamma(3 + n, x) + 8\,x\,\Gamma(4 + n, x) - 3\,\Gamma(5 + n, x)}{16\,x\,\Gamma(4 + n)}\,,
\nonumber \\
a_4 =&\ -\frac{x^6\,\Gamma(1 + n, x) - 9\,x^4\,\Gamma(3 + n, x) + 16\,x^3\,\Gamma(4 + n, x) - 9\,x^2\,\Gamma(5 + n, x) + \Gamma(7 + n, x)}{384\,x^3\,\Gamma(4 + n)}\,.
\end{align}

For the function $\cal F$ defined by Eq.~(\ref{eq25}) we get
\begin{align}
{\cal F}_{n+1,\lambda}(i\,\omega, r)=&\
-\frac1{\omega^2}\frac{d}{dr}\Big[ {\cal D}_{n+1,\lambda}(i\,\omega, r) - {\cal D}_{n+1,\lambda}(0, r)\Big]
=
\frac1{\omega^2}\frac{d}{dr}\bigg[\frac1{4\pi\,r}\, \biggl( e^{-\omega\, r} -1 +\sum_{i=1}^\infty (\omega\,r)^{2\,i}\,a_{2i}  \biggr)\bigg]
\nonumber \\ =&\
\frac1{4\pi\,\omega^2}\frac{d}{dr}\bigg(\frac{e^{-\omega\, r} -1}{r}\bigg)
+ \frac{1}{4\pi}\frac{d}{dr}\big[r\,a_2(x) + \omega^2\,r^3\,a_4(x) + \ldots  \big]
\nonumber \\ =&\
\frac1{4\pi\,\omega^2}\frac{d}{dr}\bigg(\frac{e^{-\omega\, r} -1}{r}\bigg)
+ \frac{1}{4\pi}\big[a_2(x) +x\,a'_2(x) + \omega^2\,r^2\,\big(3\,a_4(x) + x\,a'_4(x)\big) + \ldots  \big]
\nonumber \\ =&\
\frac{1-e^{-\omega\,r}\,(1+\omega\,r)}{4\pi\,\omega^2\,r^2}
+ \frac{1}{4\pi}\bigg[\frac{a_0}{2} + \omega^2\,r^2\,\frac{a_2}{4} + \ldots  \bigg].
\end{align}

\end{widetext}


\begin{thebibliography}{25}
\expandafter\ifx\csname natexlab\endcsname\relax\def\natexlab#1{#1}\fi
\expandafter\ifx\csname bibnamefont\endcsname\relax
  \def\bibnamefont#1{#1}\fi
\expandafter\ifx\csname bibfnamefont\endcsname\relax
  \def\bibfnamefont#1{#1}\fi
\expandafter\ifx\csname citenamefont\endcsname\relax
  \def\citenamefont#1{#1}\fi
\expandafter\ifx\csname url\endcsname\relax
  \def\url#1{\texttt{#1}}\fi
\expandafter\ifx\csname urlprefix\endcsname\relax\def\urlprefix{URL }\fi
\providecommand{\bibinfo}[2]{#2}
\providecommand{\eprint}[2][]{\url{#2}}

\bibitem[{\citenamefont{Fricke and Heilig}(2004)}]{fricke:04}
\bibinfo{author}{\bibfnamefont{G.}~\bibnamefont{Fricke}} \bibnamefont{and}
  \bibinfo{author}{\bibfnamefont{K.}~\bibnamefont{Heilig}},
  \emph{\bibinfo{title}{Nuclear Charge Radii}} (\bibinfo{publisher}{Springer},
  \bibinfo{year}{2004}).

\bibitem[{\citenamefont{Indelicato}(2013)}]{indelicato:13}
\bibinfo{author}{\bibfnamefont{P.}~\bibnamefont{Indelicato}},
  \bibinfo{journal}{Phys. Rev. A} \textbf{\bibinfo{volume}{87}},
  \bibinfo{pages}{022501} (\bibinfo{year}{2013}),
  \urlprefix\url{https://link.aps.org/doi/10.1103/PhysRevA.87.022501}.

\bibitem[{\citenamefont{Xie et~al.}(2023)\citenamefont{Xie, Naito, Li, and
  Liang}}]{xie:23}
\bibinfo{author}{\bibfnamefont{H.~H.} \bibnamefont{Xie}},
  \bibinfo{author}{\bibfnamefont{T.}~\bibnamefont{Naito}},
  \bibinfo{author}{\bibfnamefont{J.}~\bibnamefont{Li}}, \bibnamefont{and}
  \bibinfo{author}{\bibfnamefont{H.}~\bibnamefont{Liang}},
  \bibinfo{journal}{Physics Letters B} \textbf{\bibinfo{volume}{846}},
  \bibinfo{pages}{138232} (\bibinfo{year}{2023}),
  \urlprefix\url{https://www.sciencedirect.com/science/article/pii/S037026932300566X}.

\bibitem[{\citenamefont{Yoshimura et~al.}(2025)\citenamefont{Yoshimura,
  Yasunaga, Jido, Yamagata-Sekihara, and Hirenzaki}}]{yoshimura:25}
\bibinfo{author}{\bibfnamefont{K.}~\bibnamefont{Yoshimura}},
  \bibinfo{author}{\bibfnamefont{S.}~\bibnamefont{Yasunaga}},
  \bibinfo{author}{\bibfnamefont{D.}~\bibnamefont{Jido}},
  \bibinfo{author}{\bibfnamefont{J.}~\bibnamefont{Yamagata-Sekihara}},
  \bibnamefont{and}
  \bibinfo{author}{\bibfnamefont{S.}~\bibnamefont{Hirenzaki}},
  \bibinfo{journal}{Progress of Theoretical and Experimental Physics}
  \textbf{\bibinfo{volume}{2025}}, \bibinfo{pages}{113D02}
  (\bibinfo{year}{2025}), \urlprefix\url{https://doi.org/10.1093/ptep/ptaf140}.

\bibitem[{\citenamefont{Liborio et~al.}(2026)\citenamefont{Liborio, Kumar,
  Devadasan, Jones, Plummer, Hillier, and Bartok}}]{liborio:26}
\bibinfo{author}{\bibfnamefont{L.}~\bibnamefont{Liborio}},
  \bibinfo{author}{\bibfnamefont{M.}~\bibnamefont{Kumar}},
  \bibinfo{author}{\bibfnamefont{S.}~\bibnamefont{Devadasan}},
  \bibinfo{author}{\bibfnamefont{P.}~\bibnamefont{Jones}},
  \bibinfo{author}{\bibfnamefont{M.}~\bibnamefont{Plummer}},
  \bibinfo{author}{\bibfnamefont{A.}~\bibnamefont{Hillier}}, \bibnamefont{and}
  \bibinfo{author}{\bibfnamefont{A.}~\bibnamefont{Bartok}},
  \emph{\bibinfo{title}{Mudirac 1.3.0: A sustainable software tool for
  calculating ground state nuclear properties using muonic x-ray measurements}}
  (\bibinfo{year}{2026}), \eprint{2605.00554},
  \urlprefix\url{https://arxiv.org/abs/2605.00554}.

\bibitem[{\citenamefont{Shabaev}(1985)}]{shabaev:85}
\bibinfo{author}{\bibfnamefont{V.~M.} \bibnamefont{Shabaev}},
  \bibinfo{journal}{Teor. Mat. Fiz.} \textbf{\bibinfo{volume}{63}},
  \bibinfo{pages}{394 } (\bibinfo{year}{1985}), \bibinfo{note}{[Theor. Math.
  Phys. {\bf 63}, 588 -- 596 (1985)]},
  \urlprefix\url{https://doi.org/10.1007/BF01017505}.

\bibitem[{\citenamefont{Shabaev}(1988)}]{shabaev:88}
\bibinfo{author}{\bibfnamefont{V.~M.} \bibnamefont{Shabaev}},
  \bibinfo{journal}{Yad. Fiz.} \textbf{\bibinfo{volume}{47}},
  \bibinfo{pages}{107 } (\bibinfo{year}{1988}), \bibinfo{note}{[Sov. J. Nucl.
  Phys. {\bf 47} (1), 69 -- 72 (January 1988)]},
  \urlprefix\url{https://www.osti.gov/biblio/7021630}.

\bibitem[{\citenamefont{Shabaev}(1998)}]{shabaev:98:rectheo}
\bibinfo{author}{\bibfnamefont{V.~M.} \bibnamefont{Shabaev}},
  \bibinfo{journal}{Phys. Rev. A} \textbf{\bibinfo{volume}{57}},
  \bibinfo{pages}{59 } (\bibinfo{year}{1998}),
  \urlprefix\url{https://link.aps.org/pdf/10.1103/PhysRevA.57.59}.

\bibitem[{\citenamefont{Pachucki and Grotch}(1995)}]{pachucki:95}
\bibinfo{author}{\bibfnamefont{K.}~\bibnamefont{Pachucki}} \bibnamefont{and}
  \bibinfo{author}{\bibfnamefont{H.}~\bibnamefont{Grotch}},
  \bibinfo{journal}{Phys. Rev. A} \textbf{\bibinfo{volume}{51}},
  \bibinfo{pages}{1854 } (\bibinfo{year}{1995}),
  \urlprefix\url{https://link.aps.org/doi/10.1103/PhysRevA.51.1854}.

\bibitem[{\citenamefont{Pachucki and Yerokhin}(2023)}]{pachucki:23:prl}
\bibinfo{author}{\bibfnamefont{K.}~\bibnamefont{Pachucki}} \bibnamefont{and}
  \bibinfo{author}{\bibfnamefont{V.~A.} \bibnamefont{Yerokhin}},
  \bibinfo{journal}{Phys. Rev. Lett.} \textbf{\bibinfo{volume}{130}},
  \bibinfo{pages}{053002} (\bibinfo{year}{2023}),
  \urlprefix\url{https://link.aps.org/doi/10.1103/PhysRevLett.130.053002}.

\bibitem[{\citenamefont{Pachucki and Yerokhin}(2024)}]{pachucki:24:hpqed}
\bibinfo{author}{\bibfnamefont{K.}~\bibnamefont{Pachucki}} \bibnamefont{and}
  \bibinfo{author}{\bibfnamefont{V.~A.} \bibnamefont{Yerokhin}},
  \bibinfo{journal}{Phys. Rev. A} \textbf{\bibinfo{volume}{110}},
  \bibinfo{pages}{032804} (\bibinfo{year}{2024}),
  \urlprefix\url{https://link.aps.org/doi/10.1103/PhysRevA.110.032804}.

\bibitem[{\citenamefont{Beyer et~al.}(2026)\citenamefont{Beyer, Valuev,
  Mandrykina, Sun, and Oreshkina}}]{beyer:26}
\bibinfo{author}{\bibfnamefont{K.~A.} \bibnamefont{Beyer}},
  \bibinfo{author}{\bibfnamefont{I.~A.} \bibnamefont{Valuev}},
  \bibinfo{author}{\bibfnamefont{Z.~A.} \bibnamefont{Mandrykina}},
  \bibinfo{author}{\bibfnamefont{Z.}~\bibnamefont{Sun}}, \bibnamefont{and}
  \bibinfo{author}{\bibfnamefont{N.~S.} \bibnamefont{Oreshkina}},
  \bibinfo{journal}{Communications Physics} \textbf{\bibinfo{volume}{9}},
  \bibinfo{pages}{286} (\bibinfo{year}{2026}),
  \urlprefix\url{https://www.nature.com/articles/s42005-026-02840-z}.

\bibitem[{\citenamefont{Dumbrajs}(1980)}]{dumbrajs:80}
\bibinfo{author}{\bibfnamefont{O.}~\bibnamefont{Dumbrajs}},
  \bibinfo{journal}{Physical Review C} \textbf{\bibinfo{volume}{21}},
  \bibinfo{pages}{1677} (\bibinfo{year}{1980}),
  \urlprefix\url{https://link.aps.org/doi/10.1103/PhysRevC.21.1677}.

\bibitem[{\citenamefont{Alberto et~al.}(2021)\citenamefont{Alberto, Drago,
  Mangoni, Moretti, and Pacetti}}]{alberto:21}
\bibinfo{author}{\bibfnamefont{P.}~\bibnamefont{Alberto}},
  \bibinfo{author}{\bibfnamefont{A.}~\bibnamefont{Drago}},
  \bibinfo{author}{\bibfnamefont{A.}~\bibnamefont{Mangoni}},
  \bibinfo{author}{\bibfnamefont{S.}~\bibnamefont{Moretti}}, \bibnamefont{and}
  \bibinfo{author}{\bibfnamefont{S.}~\bibnamefont{Pacetti}},
  \bibinfo{journal}{Journal of Physics G: Nuclear and Particle Physics}
  \textbf{\bibinfo{volume}{48}}, \bibinfo{pages}{085007}
  (\bibinfo{year}{2021}).

\bibitem[{\citenamefont{Pachucki et~al.}(2025)\citenamefont{Pachucki,
  Patk\'o\v{s}, and Yerokhin}}]{pachucki:25:recfns}
\bibinfo{author}{\bibfnamefont{K.}~\bibnamefont{Pachucki}},
  \bibinfo{author}{\bibfnamefont{V.}~\bibnamefont{Patk\'o\v{s}}},
  \bibnamefont{and} \bibinfo{author}{\bibfnamefont{V.~A.}
  \bibnamefont{Yerokhin}}, \bibinfo{journal}{Phys. Rev. A}
  \textbf{\bibinfo{volume}{111}}, \bibinfo{pages}{032820}
  (\bibinfo{year}{2025}),
  \urlprefix\url{https://link.aps.org/doi/10.1103/PhysRevA.111.032820}.

\bibitem[{\citenamefont{Sun et~al.}(2025)\citenamefont{Sun, Beyer, Mandrykina,
  Valuev, Keitel, and Oreshkina}}]{sun:25}
\bibinfo{author}{\bibfnamefont{Z.}~\bibnamefont{Sun}},
  \bibinfo{author}{\bibfnamefont{K.~A.} \bibnamefont{Beyer}},
  \bibinfo{author}{\bibfnamefont{Z.~A.} \bibnamefont{Mandrykina}},
  \bibinfo{author}{\bibfnamefont{I.~A.} \bibnamefont{Valuev}},
  \bibinfo{author}{\bibfnamefont{C.~H.} \bibnamefont{Keitel}},
  \bibnamefont{and} \bibinfo{author}{\bibfnamefont{N.~S.}
  \bibnamefont{Oreshkina}}, \bibinfo{journal}{Phys. Rev. Lett.}
  \textbf{\bibinfo{volume}{135}}, \bibinfo{pages}{163002}
  (\bibinfo{year}{2025}),
  \urlprefix\url{https://link.aps.org/doi/10.1103/h3xz-xdxr}.

\bibitem[{\citenamefont{Yerokhin and Ohayon}(2026)}]{yerokhin:26:rms}
\bibinfo{author}{\bibfnamefont{V.~A.} \bibnamefont{Yerokhin}} \bibnamefont{and}
  \bibinfo{author}{\bibfnamefont{B.}~\bibnamefont{Ohayon}},
  \bibinfo{journal}{Phys. Rev. A} \textbf{\bibinfo{volume}{113}},
  \bibinfo{pages}{012804} (\bibinfo{year}{2026}),
  \urlprefix\url{https://link.aps.org/doi/10.1103/71d6-w383}.

\bibitem[{\citenamefont{Artemyev et~al.}(1995)\citenamefont{Artemyev, Shabaev,
  and Yerokhin}}]{artemyev:95:pra}
\bibinfo{author}{\bibfnamefont{A.~N.} \bibnamefont{Artemyev}},
  \bibinfo{author}{\bibfnamefont{V.~M.} \bibnamefont{Shabaev}},
  \bibnamefont{and} \bibinfo{author}{\bibfnamefont{V.~A.}
  \bibnamefont{Yerokhin}}, \bibinfo{journal}{Phys. Rev. A}
  \textbf{\bibinfo{volume}{52}}, \bibinfo{pages}{1884} (\bibinfo{year}{1995}),
  \urlprefix\url{https://link.aps.org/pdf/10.1103/PhysRevA.52.1884}.

\bibitem[{\citenamefont{Pachucki}(2024)}]{pachucki:24:hfs}
\bibinfo{author}{\bibfnamefont{K.}~\bibnamefont{Pachucki}},
  \bibinfo{journal}{Phys. Rev. A} \textbf{\bibinfo{volume}{109}},
  \bibinfo{pages}{052822} (\bibinfo{year}{2024}),
  \urlprefix\url{https://link.aps.org/doi/10.1103/PhysRevA.109.052822}.

\bibitem[{\citenamefont{Shabaev and Artemyev}(1994)}]{shabaev:94:rec}
\bibinfo{author}{\bibfnamefont{V.~M.} \bibnamefont{Shabaev}} \bibnamefont{and}
  \bibinfo{author}{\bibfnamefont{A.~N.} \bibnamefont{Artemyev}},
  \bibinfo{journal}{J. Phys. B} \textbf{\bibinfo{volume}{27}},
  \bibinfo{pages}{1307 } (\bibinfo{year}{1994}).

\bibitem[{\citenamefont{Yerokhin and Oreshkina}(2023)}]{yerokhin:23:recoil}
\bibinfo{author}{\bibfnamefont{V.~A.} \bibnamefont{Yerokhin}} \bibnamefont{and}
  \bibinfo{author}{\bibfnamefont{N.~S.} \bibnamefont{Oreshkina}},
  \bibinfo{journal}{Phys. Rev. A} \textbf{\bibinfo{volume}{108}},
  \bibinfo{pages}{052824} (\bibinfo{year}{2023}),
  \urlprefix\url{https://link.aps.org/doi/10.1103/PhysRevA.108.052824}.

\bibitem[{\citenamefont{Shabaev et~al.}(2004)\citenamefont{Shabaev, Tupitsyn,
  Yerokhin, Plunien, and Soff}}]{shabaev:04:DKB}
\bibinfo{author}{\bibfnamefont{V.~M.} \bibnamefont{Shabaev}},
  \bibinfo{author}{\bibfnamefont{I.~I.} \bibnamefont{Tupitsyn}},
  \bibinfo{author}{\bibfnamefont{V.~A.} \bibnamefont{Yerokhin}},
  \bibinfo{author}{\bibfnamefont{G.}~\bibnamefont{Plunien}}, \bibnamefont{and}
  \bibinfo{author}{\bibfnamefont{G.}~\bibnamefont{Soff}},
  \bibinfo{journal}{Phys. Rev. Lett.} \textbf{\bibinfo{volume}{93}},
  \bibinfo{pages}{130405} (\bibinfo{year}{2004}),
  \urlprefix\url{https://link.aps.org/pdf/10.1103/PhysRevLett.93.130405}.

\bibitem[{\citenamefont{Yerokhin and Shabaev}(2015)}]{yerokhin:15:recprl}
\bibinfo{author}{\bibfnamefont{V.~A.} \bibnamefont{Yerokhin}} \bibnamefont{and}
  \bibinfo{author}{\bibfnamefont{V.~M.} \bibnamefont{Shabaev}},
  \bibinfo{journal}{Phys. Rev. Lett.} \textbf{\bibinfo{volume}{115}},
  \bibinfo{pages}{233002} (\bibinfo{year}{2015}),
  \urlprefix\url{http://link.aps.org/doi/10.1103/PhysRevLett.115.233002}.

\bibitem[{\citenamefont{Rathi et~al.}(2026)\citenamefont{Rathi, Valuev, Sun,
  Heines, Indelicato, Ohayon, and Oreshkina}}]{rathi:26}
\bibinfo{author}{\bibfnamefont{S.}~\bibnamefont{Rathi}},
  \bibinfo{author}{\bibfnamefont{I.~A.} \bibnamefont{Valuev}},
  \bibinfo{author}{\bibfnamefont{Z.}~\bibnamefont{Sun}},
  \bibinfo{author}{\bibfnamefont{M.}~\bibnamefont{Heines}},
  \bibinfo{author}{\bibfnamefont{P.}~\bibnamefont{Indelicato}},
  \bibinfo{author}{\bibfnamefont{B.}~\bibnamefont{Ohayon}}, \bibnamefont{and}
  \bibinfo{author}{\bibfnamefont{N.~S.} \bibnamefont{Oreshkina}},
  \emph{\bibinfo{title}{Theory framework for medium-mass muonic atoms}}
  (\bibinfo{year}{2026}), \eprint{2603.22021},
  \urlprefix\url{https://arxiv.org/abs/2603.22021}.

\bibitem[{\citenamefont{Maximon and Schrack}(1966)}]{maximon:66}
\bibinfo{author}{\bibfnamefont{L.}~\bibnamefont{Maximon}} \bibnamefont{and}
  \bibinfo{author}{\bibfnamefont{R.}~\bibnamefont{Schrack}},
  \bibinfo{journal}{Journal of Research of the National Bureau of Standards:
  Mathematics and mathematical physics. B} \textbf{\bibinfo{volume}{70}},
  \bibinfo{pages}{85} (\bibinfo{year}{1966}).

\end{thebibliography}

\end{document}